\documentclass{sigchi}

\usepackage{balance}  
\usepackage{graphics} 
\usepackage{url}      
\usepackage{color}
\usepackage{enumitem}
\usepackage{listings}
\usepackage{pxfonts}
\usepackage{times}    
\usepackage{color}
\usepackage{float}
\usepackage{comment}

\definecolor{codegreen}{rgb}{.38,0.65,.35}
\definecolor{codegray}{rgb}{0.69,0.69,0.69}
\definecolor{codepurple}{rgb}{0.58,0,0.82}
\definecolor{codeblue}{rgb}{0.3,.45,0.6}
\definecolor{codeblack}{rgb}{.08,.08,.08}

\lstdefinestyle{mystyle}{
  commentstyle=\color{codegreen},
  keywordstyle=\color{codeblack},
  numberstyle=\tiny\color{codepurple},
  stringstyle=\color{codegreen},
  basicstyle=\scriptsize\ttfamily\color{codeblack},
  breakatwhitespace=false,         
  breaklines=true,                 
  captionpos=b,                    
  keepspaces=true,                 
  numbers=none,                    
  numbersep=5pt,                  
  showspaces=false,                
  showstringspaces=false,
  showtabs=false,                  
  tabsize=2,
  escapeinside={(*@}{@*)},
}

\definecolor{light-gray}{gray}{0.95}
\definecolor{almost-black}{gray}{0.3}

\makeatletter
\def\url@leostyle{%
  \@ifundefined{selectfont}{\def\UrlFont{\sf}}{\def\UrlFont{\small\bf\ttfamily}}}
\makeatother
\def\pprw{8.5in}
\def\pprh{11in}

\usepackage[pdftex]{hyperref}
\hypersetup{
pdftitle={Augur: Mining Human Behaviors from Fiction to Power Interactive Systems},
pdfauthor={},
pdfkeywords={information extraction, fiction, crowdsourcing, data mining},
bookmarksnumbered,
pdfstartview={FitH},
colorlinks,
citecolor=black,
filecolor=black,
linkcolor=black,
urlcolor=black,
breaklinks=true,
}

\newcommand\tabhead[1]{\small\textbf{#1}}

\newenvironment{helvetica}{\fontfamily{phv}\selectfont}{\par}

\begin{document}

\title{Augur: Mining Human Behaviors from Fiction to Power \\ Interactive Systems}

\numberofauthors{1}
\author{
  \alignauthor Ethan Fast, William McGrath, Pranav Rajpurkar, Michael S. Bernstein\\
   \affaddr{Stanford University}\\
   \email{\{ethan.fast, wmcgrath, pranavsr, msb\}@cs.stanford.edu}\\
}

\maketitle

\begin{abstract}
From smart homes that prepare coffee when we wake, to phones that know not to interrupt us during important conversations, our collective visions of HCI imagine a future in which computers understand a broad range of human behaviors. Today our systems fall short of these visions, however, because this range of behaviors is too large for designers or programmers to capture manually. In this paper, we instead demonstrate it is possible to mine a broad knowledge base of human behavior by analyzing more than one billion words of modern fiction. Our resulting knowledge base, Augur, trains vector models that can predict many thousands of user activities from surrounding objects in modern contexts: for example, whether a user may be eating food, meeting with a friend, or taking a selfie. Augur uses these predictions to identify actions that people commonly take on objects in the world and estimate a user's future activities given their current situation. We demonstrate Augur-powered, activity-based systems such as a phone that silences itself when the odds of you answering it are low, and a dynamic music player that adjusts to your present activity. A field deployment of an Augur-powered wearable camera resulted in 96\% recall and 71\% precision on its unsupervised predictions of common daily activities. A second evaluation where human judges rated the system's predictions over a broad set of input images found that 94\% were rated sensible.
\end{abstract}

\keywords{
  information extraction; fiction; crowdsourcing; data mining
}

\category{H.5.2.}{Information Interfaces and Presentation}{Graphical user interfaces}

\section{Introduction}

\begin{figure}[!t]
\centering
\includegraphics[width=1.0\columnwidth]{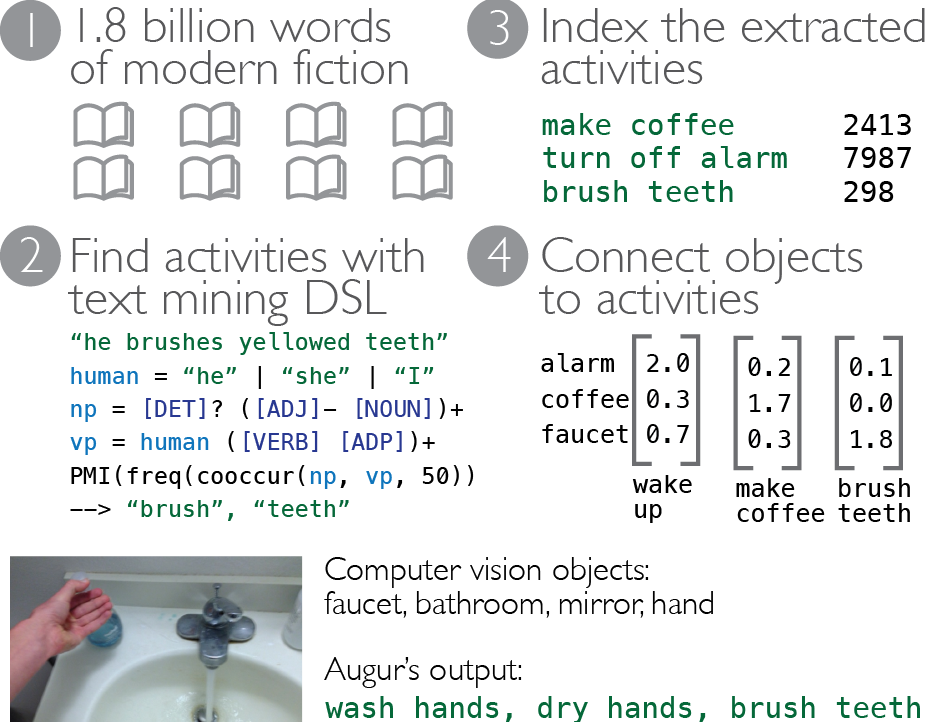}
\caption{Augur mines human activities from a large dataset of modern fiction. Its statistical associations give applications an understanding of when each activity might be appropriate.}
\label{fig:splash}
\end{figure}


Our most compelling visions of human-computer interaction depict worlds in which computers understand the breadth of human life.
Mark Weiser's first example scenario of ubiquitous computing, for instance, imagines a smart home that predicts its user may want coffee upon waking up \cite{weiser}. Apple's Knowledge Navigator similarly knows not to let the user's phone ring during a conversation \cite{knowledgenavigator}. In science fiction, technology plays us upbeat music when we are sad, adjusts our daily routines to match our goals, and alerts us when we leave the house without our wallet.
In each of these visions, computers understand the actions people take, and when.

If a broad understanding of human behavior is needed, no method yet exists that would produce it.
Today, interaction designers instead create special-case rules and single-use machine learning models. The resulting systems can, for example, teach a phone (or Knowledge Navigator) not to respond to calls during a calendar meeting. But even the most clever developer cannot encode behaviors and responses for every human activity -- we \textit{also} ignore calls while eating lunch with friends, doing focused work, or using the restroom, among many other situations. 
These original HCI visions assumed such breadth of information. 
To achieve this breadth, we need a knowledge base of human activities, the situations in which they occur, and the causal relationships between them.
Even the web and social media, serving as large datasets of human record, do not offer this information readily.

In this paper, we show it is possible to create a broad knowledge base of human behavior by text mining a large dataset of \emph{modern fiction}.
Fictional human lives provide surprisingly accurate accounts of real human activities.
While we tend to think about stories in terms of the dramatic and unusual events that shape their plots, stories are also filled with prosaic information about how we navigate and react to our everyday surroundings. 
Over many millions of words, these mundane patterns are far more common than their dramatic counterparts. Characters in modern fiction turn on the lights after entering rooms; they react to compliments by blushing; they do not answer their phones when they are in meetings. Our knowledge base, \emph{Augur} (Figure~\ref{fig:splash}), learns these associations between activities and objects by mining \emph{1.8 billion words} of modern fiction from the online writing community Wattpad.

Our main technical contribution is a vector space model for predicting human activities, powered by a domain specific language for data mining unstructured natural langage text. The domain specific language, TC, enables simple parser scripts that recognize syntactic patterns in a corpus (e.g., a verb with a human pronoun or name as its subject). TC compiles these scripts into parser combinators, which combine into more complex parsers that can, for example, run a co-occurrence analysis to identify objects that often appear near a human activity. We represent this information via a vector space model that uses smoothed co-occurrence statistics to represent each activity. For example, Augur maps \emph{eating} onto hundreds of food items and relevant tools such as cutlery, plates and napkins, and connects disconnected objects such as a fruit and store onto activities like \emph{buying groceries}. Similar parser scripts can extract subject-verb-object patterns such as ``he sips coffee'' or ``coffee spills on the table'' to understand common object behaviors. For example, Augur learns that coffee can be \textit{sipped}, and is more likely to be \textit{spilled} when another person \textit{appears}.

To enable software development under Augur, we expose three core abstractions. \emph{Activity detection} maps a set of objects onto an activity. \emph{Object affordances} return a list of actions that can be taken on or by an object. \emph{Activity prediction} uses temporal sequences of actions in fiction to predict which activities may occur next. Augur allows applications to subscribe to events on these three APIs, firing when an activity or prediction is likely. For example, a 
heads up display might register a listener onto \emph{pay money}, which fires when objects such as cash registers, bills, and credit cards are recognized by a computer vision system. The application might then react by making your bank balance available with a single tap. 
Using this API, we have created a proof-of-concept application called \textit{Soundtrack for Life}, a Google Glass application that plays music to suit your current activity.

Our evaluation tests the external validity of Augur. To examine whether the nature of fiction introduces bias into Augur's system predictions, we compare the frequency counts of its mined activities to direct human estimates, finding that 84\% of its estimates are nearly indistinguishable from human estimates. Next, we perform a field deployment of \textit{Soundtrack for Life} to collect a realistic two-hour dataset of daily activities with a human subject and then manually evaulate Augur's precision and recall. Augur demonstrated 97\% recall and 76\% precision over these daily activities. Finally, we stress-test Augur on a broader set of activities by asking external raters to evaluate its predictions on widely ranging images drawn from a sample of 50 \textit{\#dailylife} posts on Instragram. Here, raters classified 94\% of Augur's predictions as sensible, even though computer vision extracted the most semantically relevant objects in only 64\% of the images.

Our work contributes infrastructure and interfaces that draw on broad information about the associations between human interactions in the world. These interfaces can query how human behavior is conditioned by the context around the user. We demonstrate that fiction can provide deep insight into the minutae of human behavior, and present an architecture based a domain specific language for text mining that can extract this information at scale.

\section{Related Work}

Our work is inspired by techniques for mining user behavior from data. For example, query-feature graphs show how to encode the relationships between high-level descriptions of user goals and underlying features of a system \cite{qfgraphs}, even when these high-level descriptions are different from an application's domain language \cite{commandspace}. Researchers have applied these techniques to applications such as AutoCAD \cite{communitycommands} and Photoshop \cite{commandspace}, where the user's description of a domain and that domain's underlying mechanics are often disjoint. With Augur, we introduce techniques that mine real-world human activities that typically occur outside of software.

Other systems have developed powerful domain-specific support by leveraging user traces. For example, in the programming community, research systems have captured emergent practice in open source code \cite{codex}, drawn on community support for debugging computer programs \cite{helpmeout}, and modeled how developers backtrack and revise their programs \cite{backtracking}. In mobile computing, the space of user actions is small enough that it is often possible to predict upcoming actions \cite{yanglipredictions}. In design, a large dataset of real-world web pages can help guide designers to find appropriate ideas \cite{Kumar2013}. Creativity-support applications can use such data to suggest backgrounds or alternatives to the current document \cite{lee2011shadowdraw,Simon2008}. 
Augur complements these techniques by focusing on unstructured data such as text and modeling everyday life rather than behavior within the bounds of one program. 

Ubiquitous computing research and context-aware computing aim to empower interfaces to benefit from the context in which they are being used \cite{fabryq, contextaware}. Their visions motivated the creation of our knowledge base (e.g., \cite{weiser, knowledgenavigator}). Some applications have aimed to model specific activities or contexts such as jogging and cycling (e.g., \cite{ubifit}). Augur aims to augment these models with a broader understanding of human life. For example, what objects might be nearby before someone starts jogging? What activities do people perform before they decide to go jogging? Doing so could improve the design and development of many such applications.

We draw on work in natural language processing, information extraction, and computer vision to distill human activites from fiction. Prior work discusses how to extract patterns from text by parsing sentences \cite{nlpnarrative, reverb, relgram, tokens-regex}. We adapt and extend these approaches in our text mining domain-specific language, producing an alternative that is more declarative and potentially easier to inspect and reason about.
Other work in NLP and CV has shown how vector space models can extract useful patterns from text \cite{vectorspace}, or how other machine learning algorithms can generate accurate image labels \cite{cv} and classify images given a small closed set of human actions \cite{cv-act}. Augur draws on insights from these approaches to make conditional predictions over thousands of human activities.

Our research also benefits from prior work in commonsense knowledge representation. Existing databases of linguistic and commonsense knowledge provide networks of facts that computers should know about the world \cite{conceptnet}. Augur captures a set of relations that focus more deeply on human behavior and the causal relationships between human activities. We draw on forms of commonsense knowledge, like the WordNet hierarchy of synonym sets \cite{wordnet}, to more precisely extract human activities from fiction. Parts of this vocabulary may be mineable from social media, if they are of the sort that people are likely to advertise on Twitter \cite{emreactions}. We find that fiction offers a broader set of local activities.

\section{Augur}

Augur is a knowledge base that uses fiction to connect human activities to objects and their behaviors. We begin with an overview of the basic activities, objects, and object affordances in Augur, then then explain our approach to text mining and modeling.


\subsection{Human Activities}

Augur is primarily oriented around \textit{human activities}, which we learn from verb phrases that have human subjects, for example ``he opens the fridge'' or ``we turn off the lights.'' Through co-occurrence statistics that relate objects and activities, Augur can map contextual knowledge onto human behavior. 
For example, we can ask Augur for the five activities most related to the object ``facebook'' (in modern fiction, characters use social media with surprising frequency):

\vspace{.2em}
\begin{lstlisting}
activity           score       frequency
----------------------------------------
message            0.71        1456
get message        0.53        4837
chat               0.51        4417
close laptop       0.45        1480
open laptop        0.39        1042
\end{lstlisting}
\vspace{-.4em}

Here \textit{score} refers to the cosine similarity between a vector-embedded query and activities in the Augur knowledge base (we'll soon explain how we arrive at this measure). 

Like real people, fictional characters waste plenty of time \textit{messaging} or \textit{chatting} on Facebook. They also engage in activities like \textit{post}, \textit{block}, \textit{accept}, or \textit{scroll feed}.

Similarly, we can look at relations that connect multiple objects. What activities occur around a shirt and tie? Augur captures not only the obvious sartorial applications, but notices that shirts and ties often follow specific other parts of the morning routine such as \emph{take shower}:

\vspace{.2em}
\begin{lstlisting}
activity          score        frequency
----------------------------------------
wear              0.05         58685
change            0.04         56936
take shower       0.04         14358
dress             0.03         16701
slip              0.03         59965
\end{lstlisting}
\vspace{-.4em}

In total, Augur relates 54,075 human activities to 13,843 objects and locations.
While the head of the distribution contributes many observed activities  (e.g., extremely common activities like \textit{ask} or \textit{open door}), a more significant portion lie in the bulk of the tail. These less common activities, like \textit{reply to text message} or \textit{take shower}, make up much of the average fictional human's existence. Further out, as the tail diminishes, we find less frequent but still semantically interesting activities like \textit{throw out flowers} or \textit{file bankruptcy}. 

Augur associates each of its activities with many objects, even activities that appear relatively infrequently. For example, \textit{unfold letter} occurs only 203 times in our dataset, yet Augur connects it to 1072 different objects (e.g., handwriting, envelope). A more frequent activity like \textit{take picture} occurs 10,249 times, and is connected with 5,250 objects (e.g., camera, instagram). The abundance of objects in fiction allows us to make inferences for a large number of activities.

\subsection{Object Affordances}
Augur also contains knowlege about \textit{object affordances}: actions that are strongly associated with specific objects. To mine object affordances, Augur looks for subject-verb-object sentences with objects either as their subject or direct object. 
Understanding these behaviors allows Augur to reason about how humans might interact with their surroundings. For example, the ten most related affordances for a car:

\vspace{.2em}
\begin{lstlisting}
activity              score       frequency
-------------------------------------------
honk horn             0.38         243
buckle seat-belt      0.37         203
roll window           0.35         279
start engine          0.34         898
shut car-door         0.33         140
open car-door         0.33         1238
park                  0.32         3183
rev engine            0.32         113
turn on radio         0.30         523
drive home            0.26         881
\end{lstlisting}
\vspace{-.4em}

Cars undergo basic interactions like \textit{roll window} and \textit{buckle seat-belt} surprisingly often. These are relatively mundane activities, yet abundant in fiction.

Like the distribution of human activities, the distribution of objects is heavy-tailed. The head of this distribution contains objects such as phone, bag, book, and window, which all appear more than one million times. The thick ``torso'' of the distribution is made of objects such as plate, blanket, pill, and wine, which appear between 30,000 and 100,000 times. On the fringes of the distribution are more idiosyncratic objects such as kindle (the e-book reader), heroin, mouthwash, and porno, which appear between 500 and 1,500 times.

\subsection{Connections between activities}
Augur also contains information about the connections between human activities. To mine for sequential activties, we can look at extracted activities that co-occur within a small span of words. Understanding which activities occur around each other allows Augur to make predictions about what a person might do next.

For example, we can ask Augur what happens after someone orders coffee:

\vspace{.2em}
\begin{lstlisting}
activity         score       frequency
--------------------------------------
eat              0.48        49347
take order       0.40        1887
take sip         0.39        11367
take bite        0.39        6914
pay              0.36        23405
\end{lstlisting}
\vspace{-.4em}

Even fictional characters, it seems, must \textit{pay} for their orders.

Likewise, Augur can use the connections between activities to determine which activities are similar to one another. For example, we can ask for activities similar to the social media photography trend of \textit{take selfie}:

\vspace{.2em}
\begin{lstlisting}
activity             score       frequency
------------------------------------------
snap picture         0.78        1195
post picture         0.76        718
take photo           0.67        1527
upload picture       0.58        121
take picture         0.57        10249
\end{lstlisting}
\vspace{-.4em}

By looking for activities with similar object co-occurrence patterns, we can find near-synonyms.

\subsection{A data mining DSL for natural language}
Creating Augur requires methods that can extract relevant information from large-scale text and then model it. Exploring the patterns in a large corpus of text is a difficult and time consuming process. While constructing Augur, we tested many hypotheses about the best way to capture human activties. For example, we asked: what level of noun phrase complexity is best? Some complexity is useful. The pattern \textit{run to the grocery store} is more informative for our purposes than \textit{run to the store}. But too much complexity can hurt predictions. If we capture phrases like \textit{run to the closest grocery store}, our data stream becomes too sparse. Worse, when iterating on these hypotheses, even the cleanest parser code tends not to be easily reusable or interpretable.

To help us more quickly and efficiently explore our dataset, we created TC (Text Combinator), a data mining DSL for natural language. TC allows us to build parsers that capture patterns in a stream of text data, along with aggregate statistics about these patterns, such as frequency and co-occurrence counts, or the mutual information (MI) between relations. TC's scripts can be easier to understand and reuse than hand-coded parsers, and its execution can be streamed and parallelized across a large text dataset.

TC programs can model syntactic and semantic patterns to answer questions about a corpus. For example, suppose we want to figure out what kinds of verbs often affect laptops:

\vspace{.2em}
\begin{lstlisting}[language=Haskell]
laptop = [DET]? ([ADJ]+)? "laptop"
verb_phrase = [VERB] laptop-
freq(red_vp)
\end{lstlisting}
\vspace{-.4em}

Here the \textit{laptop} parser matches phrases like ``a laptop'' or ``the old broken laptop'' and returns exactly the matched phrase. The \textit{verb\_phrase} parser matches pharses like ``throw the broken laptop'' and returns just the verb in the phrase (e.g., ``throw''). The \textit{freq} aggregator keeps a count of unique tokens in the output stream of the \textit{verb\_phrase} parser. On a small portion of our corpus, we see as output:

\vspace{.2em}
\begin{lstlisting}
open        11
close        7
shut         6
restart      4
\end{lstlisting}
\vspace{-.4em}

To clarify the syntax for this example: square brackets (e.g., \lstinline{[NOUN]}) define a parser that matches on a given part of speech, quotes (e.g., \lstinline{"laptop"}) matches on an exact string, whitespace is an implicit then-combinator (e.g., \lstinline{[NOUN] [NOUN]} matches two sequential nouns), a question mark (e.g., \lstinline{[DET]?} optionally matches an article like ``a'' or ``the'', also matching on the empty string), a plus (e.g., \lstinline{[VERB]+} matches on as many verbs as appear consecutively), and a minus (e.g., \lstinline{[NOUN]-} matches on a noun but removes it from the returned match). We describe TC's full set of operators in this paper's appendix.

We wrote the compiler for TC in Python. Behind the scenes, our compiler transforms an input program into a parser combinator, instantiates the parser as a Python generator, then runs the generator to lazily parse a stream of text data. Aggregation commands (e.g., \lstinline{freq} frequency counting and \lstinline{MI} for MI calculation) are also Python generators, which we compose with a parser at compile time. Given many input files, TC also supports parallel parsing and aggregation.


\subsection{Mining activity patterns from text}
To build the Augur knowledge base, we index more than one billion words of fiction writing from 600,000 stories written by more than 500,000 writers on the Wattpad writing community\footnote{\url{http://wattpad.com}}. Wattpad is a community where amateur writers can share their stories, oriented mostly towards writers of genre fiction. Our dataset includes work from 23 of these genres, including romance, science fiction, and urban fantasy, all of which are set in the modern world. 

Before processing these stories, we normalize them using the spaCy part of speech tagger and lemmatizer\footnote{\url{https://honnibal.github.io/spaCy/})}. The tagger labels each word with its appropriate part of speech given the context of a sentence. Part of speech tagging is important for words that have multiple senses and might otherwise be ambiguous. For example, ``run'' is a noun in the phrase, ``she wants to go for a run'', but a verb in the phrase ``I run into the arms of my reviewers.'' The lemmatizer converts each word into its singular and present-tense form. For example, the plural noun ``soldiers'' can be lemmatized to the singular ``soldier'' and the past tense verb ``ran'' to the present ``runs.''

\subsubsection{Activity-Object statistics}
Activity-object statistics connect commonly co-occurring objects and human activities. These statistics will help Augur detect activities from a list of objects in a scene. We define activities as verb phrases where the subject is a human, and objects as compound noun phrases, throwing away adjectives.
To generate these edges, we run the TC script:

\vspace{.2em}
\begin{lstlisting}[language=Haskell]
human_pronoun = "he" | "she" | "i" | "we" | "they"
np = [DET]? ([ADJ]- [NOUN])+ 
vp = human_pronoun ([VERB] [ADP])+
MI(freq(co-occur(np, vp, 50)))
\end{lstlisting}
\vspace{-.4em}

For example, \textit{backpack} co-occurs with \textit{pack} 2413 times, and \textit{radio} co-occurs with \textit{singing} 7987 times. Given the scale of our data, Augur's statistics produce meaningful results by focusing just on pronoun-based sentences.

In this TC script, \lstinline{MI} (mutual information, as defined by \cite{mi-ref}) processes our final co-occurence statistics to calculate the mututal information of our relations:

\footnotesize
$$MI(A,B) = \log_{10}\left(\frac{AB*corpusSize}{A*B*span}/log_{10}{2}\right)$$
\normalsize

Where $A$ and $B$ are the frequencies of two relations, the term $AB$ is the frequency of collocation between $A$ and $B$, the term $corpusSize$ is the number of words in our corpus, and $span$ is the window size for the co-occurrence analysis.

MI describes how much one term of a co-occurrence tells us about the other. For example, if people \textit{type} with every kind of object in equal amounts, then knowing there is a computer in your room doesn't mean much about whether you are typing. However, if people type with computers far more often than anything else, then knowing there is a computer in your room tells us significant information, statistically, about what you might be doing.

\subsubsection{Object-affordance statistics}

The object-affordance statistic connects objects directly to their uses and behaviors, helping Augur understand how humans can interact with the objects in a scene. We define \textit{object affordances} as verb phrases where an object serves as either the subject or direct object of the phrase, and we again capture physical objects as compound noun phrases. 
To generate these edges, we run the TC script:

\vspace{.2em}
\begin{lstlisting}[language=Haskell]
np = [DET]? ([ADJ]- [NOUN])+ 
vp = ([VERB] [ADP])+
svo = np vp np?
MI(freq(svo))
\end{lstlisting}
\vspace{-.4em}

For example, \textit{coffee} is \textit{spilled} 229 times, and \textit{facebook} is \textit{logged into} 295 times.

\subsubsection{Activity-Activity statistics}

Activity-activity statistics count the times that an activity is followed by another activity, helping Augur make predictions about what is likely to happen next. 
To generate these statistics, we run the TC script:

\vspace{.2em}
\begin{lstlisting}[language=Haskell]
human_pronoun = "he" | "she" | "i" | "we" | "they"
vp = human_pronoun ([VERB] [ADP])+
MI(freq(skip-gram(vp,2,50)))
\end{lstlisting}
\vspace{-.4em}

Activity-activity statistics tend to be more sparse, but Augur can still uncover patterns. For example, \textit{wash hair} precedes \textit{blow dry hair} 64 times, and \textit{get text} (e.g., receive a text message) precedes \textit{text back} 34 times.

In this TC script, \lstinline{skip-gram(vp,2,50)} constructs skip-grams of length $n=2$ sequential $vp$ matches on a window size of 50. Unlike co-occurrence counts, skip-grams are order-dependent, helping Augur find potential causal relationships.

\subsection{Vector space model for retrieval}
Augur's three statistics are not enough by themselves to make useful predictions. These statistics represent pairwise relationships and only allow prediction based on a single element of context (e.g., activity predictions from a single object), ignoring any information we might learn from similar co-occurrences with other terms. For many applications it is important to have a more global view of the data.

To make these global relationships available, we embed Augur's statistics into a vector space model (VSM), allowing Augur to enhance its predictions using the signal of \textit{multiple} terms. Queries based on multiple terms narrow the scope of possibility in Augur's predictions, strengthing predictions common to many query terms, and weaking those that are not. 

VSMs encode concepts as vectors, where each dimension of the vector conveys a feature relevant to the concept. For Augur, these dimensions are defined by $MI > 0$ with Laplace smoothing (by a constant value of 10), which in practice reduces bias towards uncommon human activities \cite{vecspace-meta}. 

Augur has three VSMs. 1). \textit{Object-Activity}: each vector is a human activity and its dimensions are smoothed MI between it and every object. 2). \textit{Object-Affordance}: each vector is an affordance and its dimensions are smoothed MI between it and every object. 3). \textit{Activity-Prediction}: each vector is a  activity and its dimensions are smoothed MI between it and every other activity.

To query these VSMs, we construct a new empty vector, set the indices of the terms in the query equal to 1, then find the closest vectors in the space by measuring cosine similarity.
 
\begin{figure}[!t]
\centering
\includegraphics[width=1.0\columnwidth]{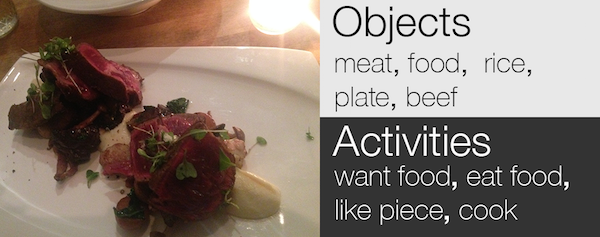}
\vspace{-0.15in}
\caption{Augur's activity detection API translates a photo into a set of likely relevant activities. For example, the user's camera might automatically photojournal the food whenever the user may be \emph{eating food}. Here, Clarifai produced the object labels.}
\label{fig:food-act}
\end{figure}

\section{Augur API and Applications}

Applications can draw from Augur's contents to identify user activities, understand the uses of objects, and make predictions about what a user might do next. To enable software development under Augur, we present these three APIs and a proof-of-concept architecture that can augment existing applications with if-this-then-that human semantics. 

We begin by introducing the three APIs individually, then demonstrate additional example applications to follow. To more robustly evaluate Augur, we have built one of these applications, \textit{Soundtrack for Life}, into Google Glass hardware.

\subsection{Identifying Activities}
\begin{figure*}[!t]
\centering
\includegraphics{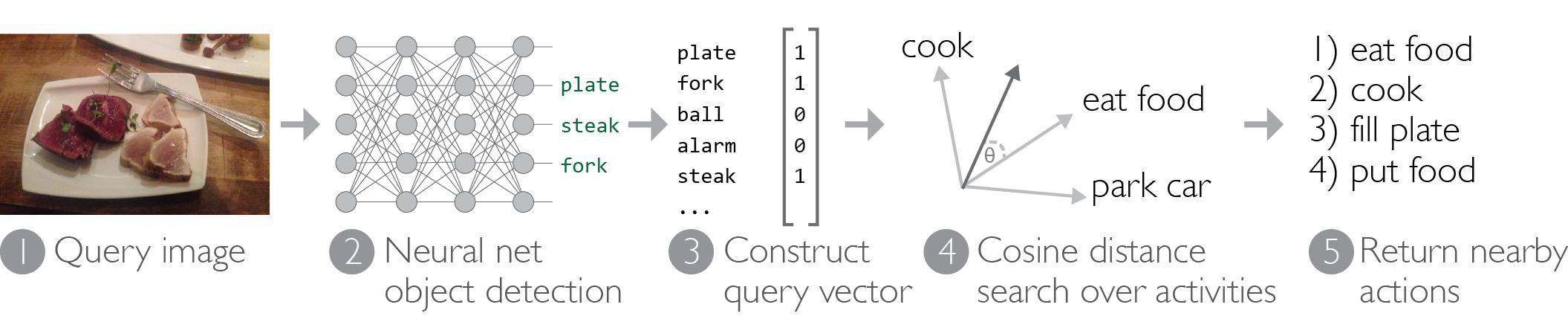}
\caption{Augur's APIs map input images through a deep learning object detector, then initializes the returned objects into a query vector. Augur then compares that vector to the vectors representing each activity in its database and returns those with lowest cosine distance.}
\label{fig:triggeroverview}
\end{figure*}

\textit{What are you currently doing?} If Augur can answer this question, applications can potentially help you with that activity, or determine how to behave given the context around you. 

Suppose a designer wants to help people stick to their diets, and she notices that people often forget to record their meals. So the designer decides to create an automatic meal photographer. She connects the user's wearable camera to a scene-level object detection computer vision algorithm such as R-CNN \cite{girshick2014rcnn}. While she could program the system to fire a photo whenever the computer vision algorithm recognizes an object class categorized as food, this would produce a large number of false positives throughout the day, and would ignore a breadth of other signals such as silverware and dining tables that might actually indicate eating.

So, the designer connects the computer vision output to Augur (Figure \ref{fig:food-act}). Instead of programming a manual set of object classes, the designer instructs Augur to fire a notification whenever the user engages in the activity \emph{eat food}. She refers to the activity using natural language, since this is what Augur has indexed from fiction:


\vspace{.2em}
\begin{lstlisting}[language=c]
image = /* capture picture from user's wearable camera */
if(augur.detect(image, "eat food"))
    augur.broadcast("take photo");

\end{lstlisting}
\vspace{-.4em}
The application takes an \lstinline{image} at regular intervals. The \lstinline{detect} function processes the latest image in that stream, pings 
a deep learning computer vision server (\url{http://www.clarifai.com/}), then runs its object results through Augur's object-activity VSM to return activity predictions. The \lstinline{broadcast} function broadcasts an object affordance request keyed on the activity \textit{take photo}: in this case, the wearable camera might respond by taking a photograph.

Now, the user sits down for dinner, and the computer vision algorithm detects a plate, steak and broccoli (Figure \ref{fig:food-act}). A query to Augur returns:

\vspace{.2em}
\begin{lstlisting}
GET /detect/plate+steak+broccoli

prediction         score       frequency
--------------------------------------
fill plate         0.39        203
put food           0.23        1046
take plate         0.15        1321
(*@\textbf{eat food}@*)           0.14        2449
set plate          0.12        740
cook               0.10        6566
\end{lstlisting}
\vspace{-.4em}

The activity \textit{eat food} appears as a strong prediction, as is (further down) the more general activity \textit{eat}. The ensemble of objects reinforce each other: when the plate, steak and broccoli are combined to form a query, eating has 1.4 times higher cosine similarity than for any of the objects individually. The camera fires, and the meal is saved for later.

\begin{figure}[!t]
\centering
\includegraphics[width=1.0\columnwidth]{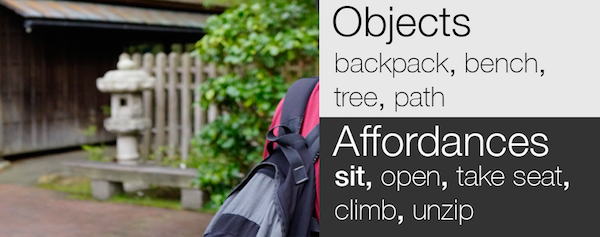}
\vspace{-0.15in}
\caption{Augur's object affordance API translates a photo into a list of possible affordances. For example, Augur could help a blind user who is wearing an intelligent camera and says they want to \emph{sit}. Here, Clarifai produced the object labels.}
\label{fig:blind}
\end{figure}

\subsection{Expanding Activites with Object Affordances}
\textit{How can you interact with your environment?} If Augur knows how you can manipulate your surroundings, it can help applications facilitate that interaction.

Object affordances can be useful for creating accessible technology. For example, suppose a blind user is wearing an intelligent camera and tells the application they want to \emph{sit} (Figure \ref{fig:blind}). Many possible objects would let this person sit down, and it would take a lot of designer effort to capture them all. Instead, using Augur's object affordance VSM, an application could scan nearby objects and find something sittable:

\vspace{.2em}
\begin{lstlisting}[language=c]
image = /* capture picture from user's wearable camera */
if(augur.affordance(image, "sit"))
  alert("sittable object ahead");
\end{lstlisting}
\vspace{-.4em}

The \lstinline{affordance} function will process the objects in the latest image, executing its block when Augur notices an object with the specified affordance. 
Now, if the user happens to be within eyeshot of a bench:

\vspace{.2em}
\begin{lstlisting}
GET /affordance/bench

prediction         score       frequency
---------------------------------------
(*@\textbf{sit}@*)                0.13        600814
take seat          0.12        24257
spot               0.11        16132
slump              0.09        8985
plop               0.07        12213

\end{lstlisting}
\vspace{-.4em}

Here the programmer didn't need to stop and think about all the scenarios or objects where a user might sit. Instead, they just stated the activity and augur figured it out.

\subsection{Predicting Future Activities}

\textit{What will you do next?} If Augur can predict your next activity, applications can react in advance to better meet your needs in that situation. Activity predictions are particularly useful for helping users avoid problematic behaviors, like forgetting their keys or spending too much money. 

In Apple's Knowledge Navigator \cite{knowledgenavigator}, the agent ignores a phone call when it knows that it would be an inappropriate time to answer. Could Augur support this? 

\vspace{.2em}
\begin{lstlisting}[language=c]
answer = augur.predict("answer call")
ignore = augur.predict("ignore call")

if(ignore > answer)
  augur.broadcast("silence phone");
else
  augur.broadcast("unsilence phone");
\end{lstlisting}
\vspace{-.4em}

The \lstinline{augur.predict} fuction makes new activity predictions based on the user's activities over the past several minutes. If the current context suggests that a user is using the restroom, for example, the prediction API will know that answering a call is an unlikely next action. When provided with an activity argument, \lstinline{augur.predict} returns a cosine similarity value reflecting the possibility of that activity happening in the near future. The activity \emph{ignore call} has less cosine similarity than \emph{answer call} for most queries to Augur. But if a query ever indicates a greater cosine similarity for \textit{ignore call}, the application can silence the phone. As before, Augur broadcasts the desired activity to any listening devices (such as the phone).

Suppose your phone rings while you are talking to your best friend about their relationship issues. Thoughtlessly, you \textit{curse}, and your phone stops ringing instantly:

\vspace{.2em}
\begin{lstlisting}
GET /predict/get call+curse

prediction       score       frequency
-------------------------------------
throw phone      0.24        3783
(*@\textbf{ignore call}@*)      0.18        567
ring             0.18        7245
answer call      0.17        4847
call back        0.17        1883
%call number      0.17        486
leave voicemail  0.17        146
\end{lstlisting}
\vspace{-.4em}

Many reactions besides cursing might also trigger \textit{ignore call}. In this case, adding \textit{curse} to the prediction mix shifts the odds between ignoring and answering significantly. Other results like \textit{throw phone} reflect the biases in fiction. We will investigate the impact of these biases in our Evaluation.

\subsection{Applications}
Augur allows developers to build situationally reactive applications across many activities and contexts. Here we present three more applications designed to illustrate the potential of its API. We have deployed one of these applications, \textit{A Soundtrack for Life}, as a Google Glass prototype.

\subsubsection{The Coffee-Aware Smart Home}

In Weiser's ubiquitous computing vision \cite{weiser}, he introduces the idea of calm computing via a scenario where a woman wakes up and her smart home asks if she wants coffee. Augur's activity prediction API can support this vision:

\vspace{.2em}
\begin{lstlisting}[language=c]
if(augur.predict("make coffee") { askAboutCoffee(); }
\end{lstlisting}
\vspace{-.4em}

Suppose that your alarm goes off, signaling to Augur that your activity is \emph{wake up}. Your smart coffeepot can start brewing when Augur predicts you want to make coffee:

\vspace{.2em}
\begin{lstlisting}
GET /predict/wake up

prediction          score       frrequency
----------------------------------------
want breakfast      0.38         852
throw blanket       0.38         728
%throw cover         0.37         1053
shake awake         0.37         774
hear shower         0.36         971
take bath           0.35         1719
(*@\textbf{make coffee}@*)         0.34         779
check clock         0.34         2408
\end{lstlisting}
\vspace{-.4em}

After people \textit{wake up} in the morning, they are likely to make coffee. They may also \textit{want breakfast}, another task a smart home might help us with.

\subsubsection{Spending Money Wisely}
We often spend more money than we have. Augur can help us maintain a greater awarness of our spending habits, and how they affect our finances. If we are reminded of our bank balence before spending money, we may be less inclined to spend it on frivolous things:

\vspace{.2em}
\begin{lstlisting}[language=c]
if(predict("pay") { 
  balance = secure_bank_query();
  speak("your balance is "+ balance); 
}
\end{lstlisting}
\vspace{-.4em}

If Augur predicts we are likely to pay for something, it will tell us how much money we have left in our account. What might triggger this prediction?

\vspace{.2em}
\begin{lstlisting}
GET /predict/enter store

prediction      score       frequency
------------------------------------
scan            0.19        5319
ring            0.19        7245
(*@\textbf{pay}@*)             0.17        23405
swipe           0.17        1800
shop            0.13        3761
%buy             0.10        32240
\end{lstlisting}
\vspace{-.4em}

For example, when you enter a store, you may be about to \textit{pay} for something. The \textit{pay} prediction also triggers on \textit{ordering} food or coffee, \textit{entering} a cafe, \textit{gambling}, and \textit{calling a taxi}.

\vspace{.2em}
\begin{lstlisting}
GET /predict/call taxi

prediction       score       frequency
------------------------------------
hail taxi        0.96        228
(*@\textbf{pay}@*)              0.96        181
take taxi        0.96        359
get taxi         0.96        368
tell address     0.95        463
get suitcase     0.82        586
\end{lstlisting}
\vspace{-.4em}

\begin{figure}[!t]
\centering
\includegraphics[width=1.0\columnwidth]{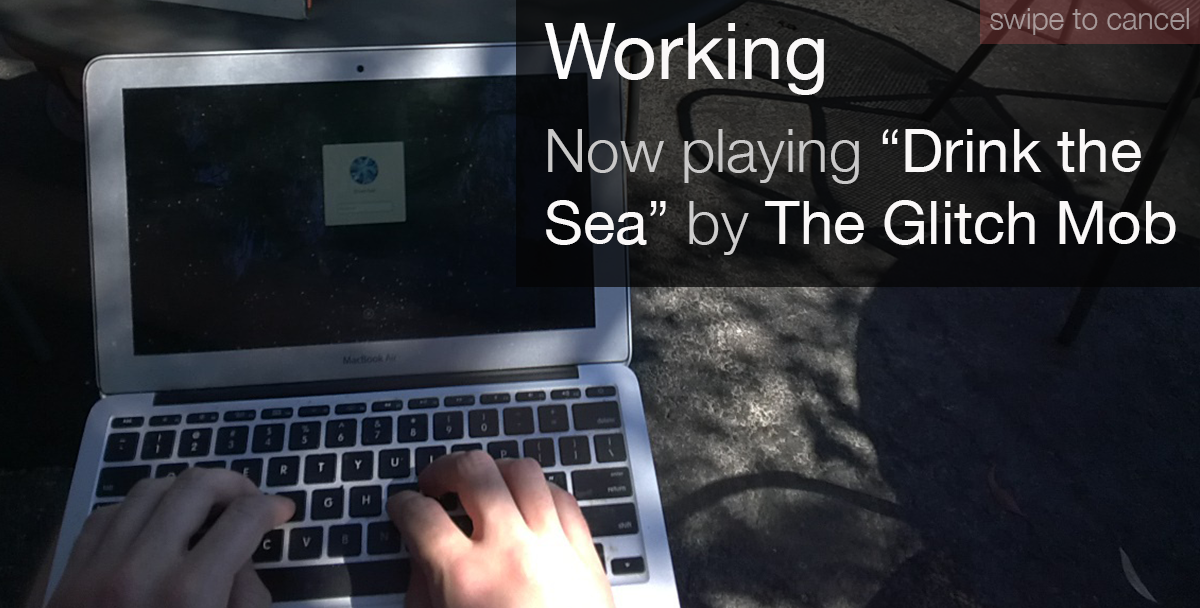}
\vspace{-0.15in}
\caption{A Soundtrack for Life is a Google Glass application that plays musicians based on the user's predicted activity, for example associating \textit{working} with The Glitch Mob.}
\label{fig:glass-test}
\end{figure}

\subsubsection{A Soundtrack for Life}
Many of life's activities are accompanied by music: you might \textit{cook} to the refined arpeggios of Vivaldi, \textit{exercise} to the dark ambivalence of St.\ Vincent, and \textit{work} to the electronic pulse of the Glitch Mob. Through an activity detection system we have built into Google Glass (Figure \ref{fig:glass-test}), Augur can arrange a soundtrack for you that suits your daily preferences. We built a physical prototype for this application as it takes advantage of the full range of activities Augur can detect. 

\vspace{.2em}
\begin{lstlisting}[language=c]
var act2music = {
    "cook": "Vivaldi",     "drive": "The Decemberists",
    "surfing": "Sea Wolf", "buy": "Atlas Genius",
    "work": "Glitch Mob",  "exercise": "St. Vincent", 
};
var act = augur.predict();
if (act in act2music){
  play(act2music[act]);
}
\end{lstlisting}
\vspace{-.4em}

For example, if you are brandishing a spoon before a pot on the stove, you are likely \textit{cooking}. Augur plays Vivaldi. 

\vspace{.2em}
\begin{lstlisting}
GET /predict/stove+pot+spoon

prediction       score       frequency
------------------------------------
(*@\textbf{cook}@*)             0.50        6566
pour             0.39        757
place            0.37        25222
stir             0.37        2610
eat              0.34        49347
\end{lstlisting}
\vspace{-.4em}

\begin{figure*}[!t]
\centering
\includegraphics[width=2.05\columnwidth]{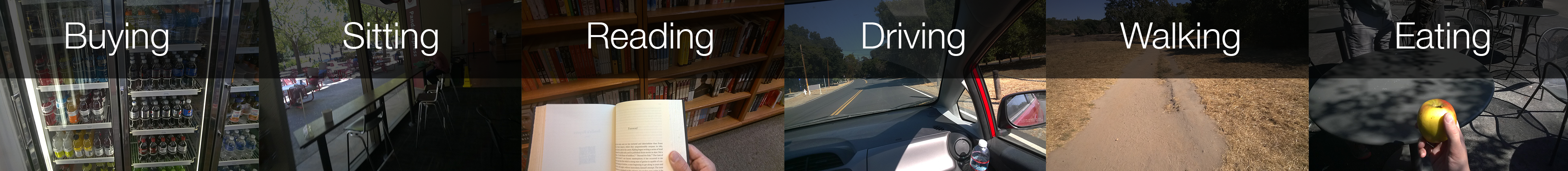}
\caption{We deployed an Augur-powered wearable camera in a field test over common daily activities, finding average rates of 96\% recall and 71\% precision for its classifications.}
\label{fig:glasstest}
\end{figure*}

\section{Evaluation}
Can fiction tell us what we need in order to endow our interactive systems with basic knowledge of human activities? In this section, we investigate this question through three studies. First, we compare Augur's activity predictions to human activity predictions in order to understand what forms of bias fiction may have introduced. Second, we test Augur's ability to detect common activities over a two-hour window of daily life. Third, to stress test Augur over a wider range of activities, we evaluate its activity predictions on a dataset of 50 images sampled from the Instagram hashtag \textit{\#dailylife}.

\subsection{Bias of Fiction}
If fiction were truly representative of our lives, we might be constantly drawing swords and kissing in the rain. Our first evaluation investigates the character and prevelance of fiction bias. We tested how closely a distribution of 1000 activities sampled from Augur's knowledge base compared against human-reported distributions. While these human-reported distributions may differ somewhat from the real world, they offer a strong sanity check for Augur's predictions.

\subsubsection{Method}
To sample the distribution of activities in Augur, we first randomly sampled 100 objects from the knowledge base. We then used Augur's activity identification API to select 10 human activities most related to each object by cosine similarity. In general, these selected activities tended to be relatively common (e.g., \textit{cross} and \textit{park} for the object ``street''). We normalized these sub-distributions such that the frequencies of their activities summed to 100. 

Next, for each object we asked five workers on Amazon Mechanical Turk to estimate the relative likelihood of its selected activities. For example, given a piano: ``Imagine a random person is around a piano 100 times. For each action in this list, estimate how many times that action would be taken. The overall counts must sum to 100.'' We asked for integer estimates because humans tend to be more accurate when estimating frequencies \cite{cogillusion}.

Finally, we computed the estimated true human distribution (ETH) as the mean distribution across the five human estimates. We compared the mean absolute error (MAE) of Augur and the individual human estimates against the ETH.

\subsubsection{Results}
Augur's MAE when compared to the ETH is 12.46\%, which means that, on average, its predictions relative to the true human distribution are off by slightly more than 12\%. The mean MAE of the individual human distributions when compared to the ETH is 6.47\%, with a standard deviation of 3.53\%. This suggests that Augur is biased, although its estimates are not far outside the variance of individual humans.

Investigating the individual distributions of activities suggests that the vast majority of Augur's prediction error is caused by a few activities in which its predictions differ radically from the humans. In fact, for 84\% of the tested activities Augur's estimate is within 4\% of the ETH. What accounts for the these few radically different estimates? 

The largest class of prediction error is caused by general activities such as \textit{look}. For example, when considering raw co-occurrence frequencies, people \textit{look} at clocks much more often than they \textit{check the time}, because \text{look} occurs far more often in general. When estimating the distribution of activities around \textit{clock}, human estimators put most of their weight on \textit{check time}, while Augur put nearly all its weight on \textit{look}. Similar mistakes involved the common but understated activities of \textit{getting into} cars or \textit{going to} stores. Human estimators favored \textit{driving} cars and \textit{shopping} at stores. 

A second and smaller class of error is caused by strong connections between dramatic events that take place more often in fiction than in real life. For example, Augur put nearly all of its prediction weight for cats on \textit{hissing} while humans distributed theirs more evenly across a cat's possible activities. In practice, we saw few of these overdramaticized instances in Augur's applications and it may be possible to use paid crowdsourcing to smooth out them out. Further, this result suggests that the ways fiction deviates from real life may be more at the macro-level of plot and situation, and less at the level of micro-behaviors. Yes, fictional characters somtimes find themselves defending their freedom in court against a murder charge. However, their actions within that courtroom do tend to mirror reality --- they don't tend to leap onto the ceiling or draw swords.

\subsection{Field test of A Soundtrack for Life}
Our second study evaluates Augur through a field test of our Glass application, \textit{A Soundtrack for Life}. We recorded a two-hour sample of one user's day, in which she walked around campus, ordered coffee, drove to a shopping center, and bought groceries, among other activities (Figure \ref{fig:glasstest}).


\subsubsection{Method}
We gave a Google Glass loaded with \textit{A Soundtrack for Life} to a volunteer and asked her, over a two hour period, to to enact the following eight activities: walk, buy, eat, read, sit, work, order, and drive. 
We then turned on the Glass, set the Soundtrack's sampling rate to 1 frame every 10 seconds, and recorded all data. 
The Soundtrack logged its predictions and images to disk.

Blind to Augur's predictions, we annotated all image frames with a set of correct activities. Frames could consist of no labeled activities, one activity, or several. For example, a subject sitting at a table filled with food might be both \textit{sitting} and \textit{eating}. We included plausible activities among this set. For example, when the subject approaches a checkout counter, we included \textit{pay} both under circumstances in which she did ultimately purchase something, and also under others in which she did not. 
Over these annotated image frames, we computed precision and recall for Augur's predictions.

\subsubsection{Results}

We find rates of 96\% recall and 71\% precision across activity predictions in the dataset (Figure \ref{tbl:soundtrack}). 
When we break up these rates by activity, Augur succeeds best at activities like \textit{walk}, \textit{buy} and \textit{read}, with precision and recall score higher than 82\%. On the other hand, we see that the activities \textit{work}, \textit{drive}, and \textit{sit} cause the majority of Augur's errors. Work is triggered by a diverse set of contextual elements. People \textit{work} at cafes or grocercy stores (for their jobs), or do construction work, or work on intellectual tasks, like writing research papers on their laptops. Our image annotations did not capture all these interpretations of work, so Augur's disagreement with our labeling is not surprising. Drive is also triggered by a large number of contexuntual elements, including broad scene descriptors like ``store'' or ``cafe,'' presumably because fictional characters often drive to these places. And \textit{sit} is problematic mostly because it is triggered by the common scene element ``tree'' (real-world people probably do this less often than fictional characters). We also observe simpler mistakes: for example, our computer vision algorithm thought the bookstore our subject visited was a restaurant, causing a large precision hit to \textit{eat}.

\begin{table}[tb]\scriptsize
\renewcommand{\arraystretch}{1.4}
\begin{helvetica}
\begin{tabular}{p{5em}@{\hspace{2em}}p{11em}@{\hspace{1em}}p{5em}@{\hspace{1em}}p{5em}}
\textbf{Activity} & \textbf{Ground Truth Frames} & \textbf{Precision} & \textbf{Recall} \\
\hline
Walk & 787 & 91\% & 99\% \\
Drive & 545 & 63\% & 100\% \\
Sit & 374 & 59\% & 86\% \\
Work & 115 & 44\% & 97\% \\
Buy & 78 & 89\% & 83\% \\
Read & 33 & 82\% & 87\% \\
Eat & 12 & 53\% & 83\% \\
\hline
\textbf{Average} &  & 71\% & 96\%
\end{tabular}
\end{helvetica}
\caption{We find average rates of 96\% recall and 71\% precision over common activities in the dataset. Here \textit{Ground Truth Frames} refers to the total number of frames labeled with each activity.}
\label{tbl:soundtrack}
\end{table}

\subsection{A stress test over \#dailylife}
Our third evaluation investigates whether a broad set of inputs to Augur would produce meaningful activity predictions.
We tested the quality of Augur's predictions on a dataset of 50 images sampled from the Instagram hashtag \textit{\#dailylife}. These images were taken in a variety of environments across the world, including homes, city streets, workplaces, restaurants, shopping malls and parks. First, we sought to measure whether Augur predicts meaningful activities given the objects in the image. Second, we compared Augur's predictions to the human activity that best describes each scene.

\subsubsection{Method}
To construct a dataset of images containing real daily activites, we sampled 50 scene images from the most recent posts to the Instagram \textit{\#dailylife} hashtag \footnote{\url{https://instagram.com/explore/tags/dailylife/}}, skipping 4 images that did not represent real scenes of people or objects, such as composite images and drawings.

We ran each image through an object detection service to produce a set of object tags, then removed all non-object tags with WordNet. 
For each group of objects, we used Augur to generate 20 activity predictions, making 1000  in total.

We used two external evaluators to independently analyze each of these predictions as to their plausibility given the input objects, and blind to the original photo. A third external evaluator decided any disagreements. High quality predictions describe a human activity that is likely given the objects in a scene: for example, using the objects \textit{street, mannequin, mirror, clothing, store} to predict the activity \textit{buy clothes}. Low quality predictions are unlikely or nonsensical, such as connecting \textit{car, street, ford, road, motor} to the activity \textit{hop}.

Next, we showed evaluators the original image and asked them to decide: 1) whether computer vision had extracted the set of objects most important to understanding the scene 2) whether one of Augur's predictions accurately described the most important activity in each scene.

\subsubsection{Results}
The evaluators rated 94\% of Augur's predictions are high quality (Table \ref{tbl:predictions}). Among the 44 that were low quality, many can be accounted for by tagging issues (e.g., ``sink'' being mistagged as a verb). The others are largely caused by relatively uncommon objects connecting to frequent and overly-abstract activities, for example the uncommon object ``tableware'' predicts ``pour cereal''.

Augur makes activity predictions that accurately describe 82\% of the images, despite the fact that CV extracted the most important objects in only 62\%. Augur's knowledge base is able to compensate for some noise in the neural net: across those images with good CV extraction, Augur succeeded at correctly predicting the most relevant activity on 94\%.

\begin{table}[tb]\footnotesize
\renewcommand{\arraystretch}{1.3}
\begin{tabular}{p{10em}@{\hspace{2em}}p{4em}@{\hspace{2em}}p{8em}}
\tabhead{Quality} & \tabhead{Samples} & \tabhead{Percent Success} \\
\hline
Augur VSM predictions & 1000 & 94\% \\
Augur VSM scene recall & 50 & 82\% \\
Computer vision object detection & 50 & 62\% \\
\end{tabular}
\caption{As rated by external experts, the majority of Augur's predictions are high-quality.}
\label{tbl:predictions}
\end{table}

\section{Discussion}
Augur's design presents a set of opportunities and limitations, many of which we hope to address in future work.

First, we acknowledge that data-driven approaches are not panaceas. Just because a pattern appears in data does not mean that it is interpretable. For example, ``boyfriend is responsible'' is a statistical pattern in our text, but it isn't necessarily useful. Life is full of uninterpretable correlations, and developers using Augur should be careful not to trigger unusual behaviors with such results. A crowdsourcing layer that verifies Augur's predictions in a specific topic area may help filter out any confusing artifacts.

Similarly, while fiction allows us to learn about an enormous and diverse set of activities, in some cases it may present a vocabulary that is too open ended. Activities may have similar meanings, or overly broad ones (like \textit{work} in our evaluation). How does a user know which to use? In our testing, we have found that choice of phrase is often unimportant. For example, the cosine similarity between \textit{hail taxi} and \textit{call taxi} is 0.97, which means any trigger for one is in practice equivalent to the other (or \textit{take taxi} or \textit{get taxi}). In this sense a large vocabulary is actively helpful. However, for other activities choice of phrase does matter, and to identify and collpase these activities, we again see potential for the refinement of Augur's model through crowdsourcing. 

In the process of pursuing this research, we found ourselves in many data mining dead ends. Human behavior is complex, and natural language is complex. Our initial efforts included heavier-handed integration with WordNet to identify object classes such as locations and peoples' names; unfortunately, ``Virginia'' is both. This results in many false positives. Likewise, activity prediction requires an order of magnitude more data to train than the other APIs given the $N^2$ nature of its skip-grams. Our initial result was that very few scenarios lent themselves to accurate activity prediction. Our solution was to simplify our model significantly (e.g., looking at only pronouns) and gather ten times the raw data from Wattpad. In this case, more data beat more modeling intelligence.

More broadly, Augur suggests a reinterpretation of our role as designers. Until now, the designer's goal in interactive systems has been to articulate the user's goals, then fashion an interface specifically to support those goals. Augur proposes a kind of ``open-space design'' where the behaviors may be left open to the users to populate, and the designer's goal is to design reactions that enable each of these goals. To support such an open-ended design methdology, we see promise in Augur's natural language descriptions. Activities such as ``sit down'', ``order dessert'' and ``go to the movies'' are not complex activity codes but human-language descriptions. We speculate that each of Augur's activities could become a command. Suppose any device in a home could respond to a request to ``turn down the lights''. Today, Siri has tens of commands; Augur has potentially thousands.

\section{Conclusion}
Interactive systems find themselves in a double bind: they cannot sense the environment well enough to build a broad model of human behavior, and they cannot support the breadth of human needs without that model. Augur proposes that fiction --- an ostensibly false record of humankind --- provides enough information to break this stalemate. Augur captures behaviors across a broad range of activities, from drinking coffee, to starting a car, to going out to a movie. These behaviors often represent concepts that designers may have never thought to hand-author.

Moving forward, we will integrate this information as a prior into the kinds of activity trackers and machine learning models that developers already use. We aim to develop a broader suite of Augur applications and test them in the wild.

\section{Acknowledgments}
Special thanks to our reviewers and colleagues at Stanford for their helpful feedback.
This work is supported by a NSF Graduate Fellowship.


\bibliographystyle{acm-sigchi}
\bibliography{ref}

\end{document}